# Curious, Critical Thinker, Empathetic, and Ethically Responsible: Essential Soft Skills for Data Scientists in Software Engineering

Matheus de Morais Leça
University of Calgary
Calgary, AB, Canada
matheus.demoraisleca@ucalgary.ca

Ronnie de Souza Santos
University of Calgary
Calgary, AB, Canada
ronnie.desouzasantos@ucalgary.ca

*Abstract*—**Background.** As artificial intelligence and AI-powered systems continue to grow, the role of data scientists has become essential in software development environments. Data scientists face challenges related to managing large volumes of data and addressing the societal impacts of AI algorithms, which require a broad range of soft skills. *Goal.* This study aims to identify the key soft skills that data scientists need when working on AI-powered projects, with a particular focus on addressing biases that affect society. *Method.* We conducted a thematic analysis of 87 job postings on LinkedIn and 11 interviews with industry practitioners. The job postings came from companies in 12 countries and covered various experience levels. The interviews featured professionals from diverse backgrounds, including different genders, ethnicities, and sexual orientations, who worked with clients from South America, North America, and Europe. *Results.* While data scientists share many skills with other software practitioners—such as those related to coordination, engineering, and management—there is a growing emphasis on innovation and social responsibility. These include soft skills like curiosity, critical thinking, empathy, and ethical awareness, which are essential for addressing the ethical and societal implications of AI. *Conclusion.* Our findings indicate that data scientists working on AI-powered projects require not only technical expertise but also a solid foundation in soft skills that enable them to build AI systems responsibly, with fairness and inclusivity. These insights have important implications for recruitment and training within software companies and for ensuring the long-term success of AI-powered systems and their broader societal impact.

## I. INTRODUCTION

In an increasingly competitive landscape, software companies must address the needs of their consumers in the best way possible. For a company to be successful, solutions must incorporate technological advancements and be thoughtfully designed to reflect the desires and challenges of the target audience [1]. Additionally, software products must reflect the diversity of users in society, ensuring inclusivity and fairness in their design and functionality [2]. To accomplish this, companies need teams composed of highly skilled technical professionals [3]. However, these workers must also possess strong soft skills to effectively address the challenges of developing diverse and user-centered solutions [4].

Soft skills are essential for maximizing the effectiveness of hard skills [5], [6], particularly when collaboration and communication are key. These skills are needed when working with others [7] or developing solutions that cater to the unique needs of diverse users [8]. They reflect a person's ability to understand and engage effectively with colleagues and stakeholders. Soft skills become especially important in large-scale projects, where successful collaboration is vital for overcoming complex challenges [9], [10].

With the development landscape rapidly changing due to advancements in technologies like artificial intelligence (AI) [11], the way systems are built, and problems are solved is changing [11]. This evolution requires developers to not only adopt new approaches but also expand their skill sets [12]. For instance, beyond technical coding proficiency, developers working with AI-powered systems must consider the broader implications these technologies bring to society [13], [14], because if not thoughtfully designed, these systems can lead to unintended and harmful consequences [15]. If not designed properly, they can be used with malicious intent or be biased against minorities. To avoid this, the teams must promote ethical awareness to ensure that these solutions ensure transparency, privacy, and accountability [16]–[18].

Today, the demand for data is greater than ever [19]. Often referred to as the "oil" of the digital era [20], data plays a crucial role in shaping technological solutions [21]. Examples of its importance include recommendation systems in streaming services [22], algorithms on social media platforms [23], and decision-making systems, all of which rely on vast amounts of data to deliver optimal outcomes. This demand is expected to increase even further since the rise of Deep Learning models, which require extensive training data [24]. In this context, the role of a data scientist has become indispensable within software development teams, as these professionals are responsible for gathering, cleaning, and analyzing large datasets [25], while building and fine-tuning machine learning models to extract meaningful insights [26].

Like any other software professional on the team, data scientists need both hard and soft skills to succeed in their roles and deliver solutions that meet the diverse demands of society

[27]. However, while there is substantial research on the soft skills needed by various software practitioners [27]–[30], there is limited knowledge about the specific soft skills required for data scientists. To address this gap, this paper focuses on identifying the soft skills essential for data scientists working on AI-powered systems, such as those involving machine learning models. In particular, we focus on answering the following research question:

> ***Research Question***: *What key soft skills are necessary for data scientists in the development of AI systems?*

Following this introduction, this paper is structured as follows. Section 2 reviews existing studies on soft skills in software development and related works. Section 3 details the methodology used in this study. Section 4 presents our findings, followed by a discussion of these results, their implications, and the study's limitations in Section 5. Finally, Section 6 concludes with a summary of the contributions of this research and suggests potential avenues for future work.

## II. BACKGROUND

In software engineering, as in many other fields, professionals need a range of skills to succeed in the workforce. These skills are typically categorized into two types: hard skills, which refer to the technical abilities required to perform specific tasks [6], and soft skills, which involve non-technical abilities such as intrapersonal and interpersonal capabilities that enhance performance in various situations [31]. Hard skills in software engineering are often straightforward and primarily focus on the main task a professional perform, e.g., coding proficiency [32]. However, soft skills can vary across different tasks and are sometimes difficult to define, depending on the context [33].

### A. Soft Skills in Software Development

Although soft skills are often seen as essential in software development, their role in enhancing practitioner performance is still not fully understood [34]. These skills can significantly improve an individual's performance across various tasks, yet, like hard skills, they are typically acquired rather than innate [35]. Therefore, for a software company to effectively incorporate these qualities into its teams, it must not only hire individuals with a strong grasp of these skills but also create an environment that nurtures and encourages their development in the workplace [36].

Given that software development is inherently a creative field, soft skills related to innovation—such as problem-solving and analytical thinking—are highly valued [37]. These abilities allow developers to tackle difficult challenges, whether it is meeting specific project requirements or diagnosing and resolving complex bugs [38]. For instance, problem-solving skills enable developers to break down complex problems into manageable tasks, identify the root causes of issues, and devise solutions that are both practical and scalable. Similarly, analytical thinking helps developers evaluate multiple solutions, assess risks, and choose the best course of action, especially when working with intricate systems or adapting to evolving project requirements. In an industry where unpredictability is common, these skills are essential for navigating both technical complexities and changing client needs, ensuring that projects stay on track and deliver high-quality outcomes [28].

Employees who excel in these areas are more likely to thrive in the software industry [28], in particular, because collaboration is a fundamental part of software development. Whether through teamwork or interactions with stakeholders, communication and coordination skills are critical [39]. Effective communication improves productivity [40] and ensures that project goals align with customer expectations, which is key to the successful delivery of software products [41]. Furthermore, the ability to explain technical concepts to non-technical team members and clients is essential for bridging knowledge gaps and avoiding misunderstandings, system errors, and budget overruns [42].

### B. Soft Skills for Specific Tasks in Software Engineering

In software development, professionals from various specializations collaborate to contribute their unique expertise, each playing an important role in building the software. Whether the people involved are project managers, programmers, software engineers, or testers, every individual brings a distinct set of skills to the table. These roles not only require technical expertise but also demand a range of soft skills to ensure that the team functions cohesively, as these professionals must work together to solve complex problems, meet user needs, and deliver a high-quality product [43].

For instance, in software teams, managers interact with people regularly, selecting team members based on their skills and assigning tasks that match their strengths. To coordinate the team effectively and keep everyone working towards a common goal, managers must have strong communication and organizational abilities [44]. Programmers and software engineers rely heavily on analytical thinking to turn user requirements into practical solutions [34]. They also need to anticipate challenges users might face when interacting with the product, considering different perspectives to create more inclusive and accessible software. Testers also need strong problem-solving skills to identify and resolve bugs effectively [45]. Their ability to find the root causes of issues and ensure the software works as intended is important for maintaining the overall quality and reliability of the final product.

As more professionals, such as UX designers, data scientists, and security specialists, become part of the software development process, they, too, need to develop soft skills that complement their technical abilities. Soft skills across all roles help create a well-rounded team where everyone contributes not only their technical knowledge but also enhances the team's overall performance through collaboration, adaptability, and user-centered thinking [28], [39].

## III. METHOD

To investigate the soft skills necessary for data science professionals who work on software projects, particularly those focused on the development of AI-powered systems, we employed a qualitative methodology [46] that utilized a dual perspective in data collection. Using this approach, we integrated insights from job postings and practitioners, which supported our exploration of the subject. By combining these perspectives, we ensured a thorough understanding of the essential soft competencies valued in data science and AI development.

First, to capture the industry's perspective on this theme, we conducted a systematic collection of job postings to examine trends and expectations regarding the soft skills that employers prioritize for data science roles. This analysis offered a broad overview of organizational demands and highlighted the key competencies deemed essential for success in the field. Simultaneously, we gathered insights from data scientists working in the software industry through in-depth interviews. By exploring their experiences, we gained a detailed understanding of how these soft skills are applied in real-world contexts and valued by professionals in the field.

By integrating these two perspectives, we leveraged the complementary strengths of different sources while mitigating their limitations [47], [48]. Specifically, the analysis of job postings revealed trends regarding the soft skills desired for data science roles but lacked depth in terms of practical application. In contrast, insights from professionals provided rich, detailed information about how these soft skills are applied in daily activities but may not fully reflect broader industry expectations. By combining both sources, we gained a more comprehensive understanding of the essential soft skills in data science, as well as newly emerging skills in this context. Next, we provide a detailed explanation of our data collection and analysis methods.

### A. LinkedIn Data Collection

We conducted a systematic collection of data from job postings on the LinkedIn platform to pinpoint the most required soft skills from a market perspective. By utilizing one of the most prominent platforms for job postings, we were able to gather relevant, meaningful, and precise information about the soft skills that employers expect from professionals in the job market.

In this process, we accessed LinkedIn and searched for job postings using the keywords *Data Scientist*, *Soft Skills*, and *Software*. Our target locations included 12 countries: China, Japan, the United States, Canada, Mexico, Brazil, Germany, the United Kingdom, Egypt, South Africa, Australia, and New Zealand. These countries were selected to ensure geographical diversity by choosing the two largest economies from each region, based on gross domestic product (GDP), from each central region: Asia, North America, South America, Europe, Africa, and Oceania [49].

The Python script retrieved over 1,600 job postings, but many were outside the scope of this research, e.g., job postings for other positions in software companies. To refine the data, we filtered the job postings to retain only those whose job titles contained one of the following terms: *Data Scientist*, *Data Analyst*, *Machine Learning*, *Data Science*, *Data Engineer*, and *Artificial Intelligence*. After applying this filter, we narrowed the dataset to 87 job postings, all related to the specific roles in the data science and AI fields.

A considerable number of job postings were excluded (e.g., from 1,600 to 87 job postings) because the search terms were applied to the full content of the job postings rather than being limited to the job titles. As a result, many irrelevant postings were initially captured, which diluted the dataset. To address this, we refined the filtering process by focusing specifically on keywords within the job titles. This adjustment ensured that the remaining job postings were directly relevant to our research on the soft skills required in data science roles, improving the overall quality and accuracy of our dataset.

### B. Interviews

To capture the perspectives of professionals involved in data science projects where soft skills play a key role, we conducted semi-structured interviews to explore their experiences. Between June 1 and July 5, 2024, we interviewed 11 professionals online using a pre-established interview script. The script covered key topics such as their work on AI-powered projects, their role within the team, their main activities, how they identify and mitigate biases in these systems, and how they interact with other team members. Additionally, we specifically asked participants to describe the soft skills they consider essential when working on the development of AI systems. The interviews lasted between 23 and 42 minutes, resulting in approximately 5 hours of audio and more than 450 pages of transcripts.

To recruit participants, we initially sent an open invitation to professionals working on these projects, asking those interested in scheduling interviews, following a convenience sampling approach [50]. After each interview, we encouraged participants to recommend colleagues who might be interested in joining the study using snowball sampling [50]. To further refine our sample, we employed theoretical sampling [51], reaching out directly to individuals who we believed could provide valuable insights into specific phenomena we were still investigating. This included targeting professionals with diverse experience levels or demographic backgrounds to deepen our understanding.

The final sample included 11 professionals actively engaged in developing AI-powered systems, such as deep learning neural networks, prediction models, large language models (LLMs), and computer vision systems for facial recognition. These professionals came from diverse backgrounds, including variations in gender, ethnicity, sexual orientation, and socioeconomic status. We prioritized this diversity in our sample due to ongoing discussions about the challenges AI systems face related to bias and fairness, as well as the recognition that diversity and inclusion are key to developing effective solutions. Additionally, the individuals in our sample worked on projects for clients from South America, North America, and Europe, providing geographic diversity in their software engineering practices. This approach ensured that our analysis captured a broad range of perspectives and experiences in the development of AI-powered systems.

*C. Data Analysis*

To identify the soft skills sought in data science professionals, we conducted a thematic analysis, a systematic method used to identify, analyze, and report patterns, or themes, within qualitative data [52]. The thematic analysis involves coding the data to highlight key concepts, grouping similar codes into broader themes, and interpreting the significance of these themes in relation to the research question. This flexible yet structured approach allows researchers to uncover both explicit references and implicit patterns within the data. In this study, thematic analysis enabled us to explore both explicit and implicit mentions of soft skills across two key data sources: the collected job postings and the transcripts from our semi-structured interviews with data science professionals.

Our analysis began by thoroughly reviewing the job postings and interview transcripts to understand the context in which soft skills were mentioned. To enhance the efficiency and accuracy of our understanding and the first step of the coding process, i.e., open coding, we employed an advanced natural language processing model from the ChatGPT platform. We utilized prompt engineering—a systematic approach to designing precise and structured inputs or queries that guide generative language models to produce accurate, context-specific outputs [53]. Prompt engineering involves iteratively refining the input instructions given to the model so that it performs desired tasks effectively. This process requires attention to the structure, phrasing, and context of the prompts to ensure optimal results [53].

In our case, prompt engineering was essential to explore the vast amount of collected data and to ensure that the model focused only on the relevant sections of the data and captured explicit mentions of soft skills. Several iterations of prompt refinement were necessary to achieve validated results and ensure that we could trust the initial coding classification performed by the tool. For example, our initial prompts caused the model to generate "hallucinations," creating examples of soft skills not present in the data. In our second iteration, we focused solely on explicit mentions, but this approach missed many relevant citations from the job postings. To address this, we initially supplied the model with a predefined list of soft skills, which limited its focus to those specific terms, missing broader relevant data. We then broadened the model's scope by using a general definition of soft skills, which allowed it to identify more references but also led to irrelevant searches and some empty or inaccurate results. Ultimately, we refined our approach by focusing the model on job descriptions within the postings and specifying it to exclude empty rows, enhancing the accuracy and relevance of the results.

After these refinements, the final prompt (shown below) allowed the model to accurately extract relevant mentions of soft skills across the data source, creating a comprehensive list of codes that we used to group into categories. We applied the same process to the interview transcriptions, focusing only on the interviewees' speech in the transcripts. The final prompt (shown below), used to collect relevant mentions of soft skills, followed a similar approach for consistency.

---

*Prompt – Job Postings.*
**Instruction**: You are a highly specialized Data Analyst focused on identifying soft skills from text. Your task is to analyze the "job description" column in each row of the dataset. For every row, follow these steps: i) Identify explicitly mentioned soft skills based on the provided definition. Extract the sentence or phrase only from the "job description" column that supports each identified soft skill; ii) Soft Skill Definition: The combination of the abilities, attitudes, habits, and personality traits that allow people to perform better in the workplace, complementing the technical skills [...] and influencing the way they behave and interact with others.
**Requirements**: Only include rows where soft skills are explicitly mentioned in the "job description" column. The Supporting Text must be a direct quote from the "job description" column. Exclude rows where no soft skills are found.
**Output**: Provide the results in a structured table with three columns: i) Job ID: The unique identifier for each job description; ii) Soft Skill: The identified soft skill; iii) Supporting Text: The exact quote from the job description column that supports the soft skill.
**Example**: For the soft skill "Communication": Supporting Text: "Strong communication skills are required to interact with clients and team members..."
**Final Output Formatting**: Only include rows where a soft skill has been identified. The table must contain three columns: Job ID, Soft Skill, and Supporting Text (all quotes must come from the job description column). Ensure the output is consistently structured and ready for Excel export.

---

*Prompt – Interviews.*
**Instruction**: You are a highly specialized Data Analyst focused on identifying soft skills from text. Your task is to analyze the text in the provided Word file and extract all explicitly mentioned soft skills from the interviewees' responses.
Soft Skill Definition: The combination of the abilities, attitudes, habits, and personality traits that allow people to perform better in the workplace, complementing the technical skills [...] and influencing the way they behave and interact with others.
**Output Requirements**: Save the results in an Excel file with the following format: i) Column 'Participant ID': Where the ID corresponds to the interview from which the text is taken. ii) Column 'Soft Skill': The name of the soft skill (e.g., 'Communication'). iii) Column 'Extracted Text': The exact text where the soft skill is mentioned (e.g., 'I believe that to be a good data scientist you need to have excellent communication....').

---

This approach ensured that the extraction of soft skills from multiple sources was both precise and contextually relevant while significantly reducing manual effort and minimizing the potential human bias inherent in traditional coding methods. By leveraging automation, we were able to streamline the process and focus more on interpretation rather than initial data handling. However, to improve the validity of the automated extraction, we performed a manual review on a randomly selected sample comprising 20% of the extracted data. This step aimed to verify the accuracy of the automated coding and identify any potential false positives or misclassifications. The manual review confirmed that all retrieved data was accurate, and no false positives were detected during this process.

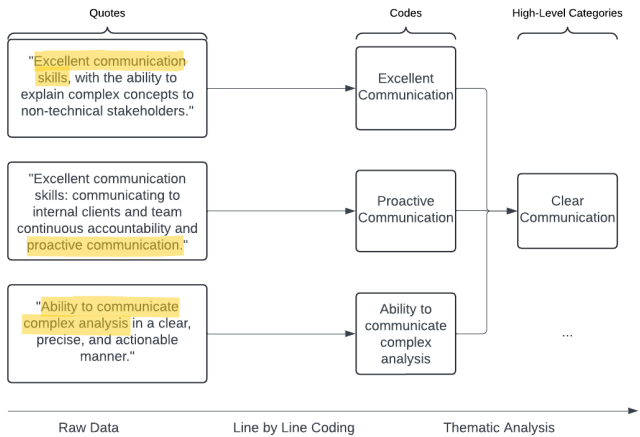

Figure 1. Coding Process

We then proceeded to the next stages of the thematic analysis by grouping similar codes into high-level categories. This step involved systematically organizing the initial codes into broader, more abstract themes, enabling us to identify patterns and relationships that linked the various soft skills together. By carefully examining how these categories interacted and overlapped, we were able to highlight underlying connections and group-related skills under comprehensive themes. As we progressed, the analysis evolved to the point where we could articulate the central narrative emerging from our data. This narrative outlined important soft skills necessary for data scientists to effectively contribute to software projects, particularly those involving complex AI systems.

### D. Ethics

This study was conducted in strict adherence to established ethical standards for research involving human subjects and data collection. All job postings were sourced exclusively from publicly available information on the LinkedIn website, ensuring that no proprietary or confidential data was utilized without authorization. The interview data were obtained from a separate survey in which participants were fully informed about the nature of the research, including any potential risks associated with their involvement. Informed consent was obtained from all participants prior to their involvement. We emphasized their rights to confidentiality and the option to withdraw from the study at any point without any repercussions. To ensure privacy, all sensitive information has been securely stored and is accessible only to authorized members of the research team. No personally identifiable information is disclosed in this paper, and all data have been anonymized to prevent any potential identification of individual participants.

### IV. FINDINGS

Our findings encompass the vision of industry and practitioners regarding the role of soft skills in the software development process, gathering insights on how these skills contribute to the success of projects. Some of the quotations provided below have been translated into English by the authors. While some may read awkwardly, we have aimed to transcribe and translate them as accurately as possible.

The analysis of the 87 job postings revealed several key insights. A substantial majority, 62%, were based in the United States, underscoring the country's strong demand for data scientists. Germany and the United Kingdom followed, accounting for 9% and 8% of the postings, respectively. This geographical distribution highlights a concentration of opportunities in North America and Europe, where demand for data science professionals is particularly high. Regarding experience levels, most positions were entry-level roles, making up 54% of the sample. Mid-senior level positions accounted for 32%, while associate-level roles made up 7%. The predominance of entry-level openings suggests a growing need for data science professionals at the foundational level, likely driven by organizations expanding their data science teams to support increasing data-driven initiatives. The significant proportion of mid-senior roles also indicates ongoing demand for experienced professionals who can lead projects and mentor junior staff.

In the interviews, the sample represented a range of data scientists working on different types of AI-powered system projects. The final sample was diverse, with 36% of participants identifying as non-male, including one non-binary software engineer. Additionally, 18% of participants were non-white, including individuals of African ancestry, and 18% identified as LGBTQIA+, including those identifying as gay, lesbian, and queer. Moreover, 9% of the sample were neurodivergent, including individuals diagnosed with ADHD. The sample was also highly experienced, with 54% having more than five years of experience in software development. This level of expertise provided a comprehensive and nuanced understanding of the challenges and perspectives involved in AI development.

### A. Data Scientist Software Skills Identified on LinkedIn Posts

We identified ten key soft skills that employers deem highly desirable in data science professionals. These soft skills were extracted from job descriptions across the collected postings, reflecting current industry demands and highlighting the competencies that companies prioritize in their hiring processes. After extracting these skills, we grouped them into five categories that made sense based on the relationships among the skills: *coordination*, *engineering*, *management*, *innovation*, and *social responsibility*. Each category reflects a core competency area required for success in data science projects. The first category, *coordination skills*, encompasses the abilities necessary for working effectively within teams and across different organizational units. These skills were referenced in 142 job postings. Next, we identified *engineering* *skills*, skills that are critical for solving the complex problems data scientists face daily, referenced 47 times in the postings. The third category, *management skills*, was emphasized 24 times and focuses on leadership and project execution. *Innovation skills* is, mentioned five times, important for fostering creativity and adaptability in the rapidly evolving field of AI and data science. Finally, *social responsibility*, mentioned three times, is related to the consideration for the impact of AI on others, e.g., the impact of AI in society.

The category of *coordination skills* includes the abilities necessary for effective teamwork and collaboration across different teams and organizational units. It covers *effective collaboration* [54], which involves working with diverse groups to achieve shared goals, making it essential for interdisciplinary projects. *Clear communication* [55] ensures that ideas, requirements, and feedback are accurately conveyed, reducing misunderstandings and keeping team members aligned. *Teamwork* [56] focuses on the ability to work cohesively in groups, prioritizing collective success over individual contributions. Additionally, fostering strong *interpersonal relationships* [57] helps build trust and a respectful environment, promoting better cooperation and enhancing team dynamics.

*Engineering skills* are vital for solving technical challenges in data science projects. This category includes *problem-solving* [54], which involves analyzing issues, identifying gaps, and finding solutions to address challenges effectively. Paired with this is *decision-making* [54], the ability to make informed choices that guide the direction of projects and ensure goals are met efficiently. These skills enable data scientists to tackle complex problems using structured, analytical approaches while adapting to the dynamic needs of data-driven environments.

*Management skills* center around leadership and the efficient execution of projects. *Leadership* [58] refers to the ability to guide and inspire teams toward achieving strategic objectives, providing direction, and ensuring everyone is aligned with the project's goals. Complementing this, *time management* [59] focuses on the ability to allocate time effectively, balancing multiple tasks and ensuring deadlines are met without sacrificing quality. Together, these skills help ensure that data science projects progress smoothly, with teams working efficiently and in sync.

The category of *innovation skills* supports creativity and adaptability, especially in the fast-changing field of AI and data science. The key skill here is *critical thinking* [60], which enables professionals to assess situations, analyze data objectively, and approach problems with fresh perspectives. By fostering creativity and encouraging out-of-the-box thinking, this skill helps data scientists develop innovative solutions that drive technological advancements and adapt to new challenges in their work.

*Social responsibility* reflects the ethical dimension of data science, focusing on the broader impact of AI on individuals and society. *Empathy* [61] is a key skill in this category, involving the ability to understand and consider the needs and emotions of others. This is important not only in team collaboration but also in creating AI systems that are user-centric and inclusive. Empathy helps ensure that data scientists develop technologies that are considerate of ethical concerns, biases, and the potential societal impacts of AI.

### B. Data Scientist Software Skills Identified in the Interviews

By analyzing the data obtained from the interview transcripts, we identified three major soft skills that data scientists perceive as valuable in their daily work activities. While these skills are closely aligned with those identified in the job postings, the interviews offer unique perspectives from the practitioners' experiences. Unlike the job postings, which were broader in terms of general skills required not just for data scientists but for any professional working on software teams, the interviews resulted in more focused skills. This is a result of professionals specifically focusing on their experiences dealing with aspects related to AI development, including challenges associated with bias in AI and its broader implications.

In this sense, in addition to *empathy*, which was previously identified in the job postings, the data scientists emphasized *ethical awareness* as another skill related to *social responsibility* necessary in their work. *Ethical awareness* [62] refers to the understanding of moral duties and obligations, particularly in managing sensitive data. This skill ensures that professionals handle data responsibly, maintaining fairness in decision-making processes. It also highlights the need for data scientists to go beyond technical skills and actively work to safeguard against biases and ethical pitfalls in AI systems, ensuring their work is both transparent and just.

In the context of *innovation skills*, curiosity emerged as a critical skill, demonstrating its role in driving creativity and adaptability. *Curiosity* [63] is an enduring personality trait that encourages individuals to explore new ideas and seek out novel solutions. For data scientists, curiosity plays a key role in staying inquisitive, investigating emerging technologies, and proactively searching for new tools and methodologies to improve AI outcomes. This is particularly important to enable professionals to stay ahead of industry trends, including regarding issues, and continuously adapt to the rapidly evolving landscape of AI and data science.

Interestingly, the professionals interviewed were less focused on commenting on general soft skills typically required for software engineers. The only broadly mentioned skill was *clear communication*, which had already been identified in the job postings. This *coordination skill* is seen as important for supporting effective teamwork and collaboration. Clear communication helps team members stay aligned, minimizes misunderstandings, and plays a key role in improving team efficiency and fostering a cohesive work environment, especially when several aspects of software fairness still lack clear definitions.

## C. Software Skills Required From Data Scientists in the Software Industry

In our approach, we combined two complementary perspectives on the key soft skills needed by data scientists: those of employers (reflected in job postings) and those of practitioners (captured through interviews). This dual perspective allowed us to build a robust understanding of the skills required in the field. However, it is important to note that the findings from the job postings and interviews diverge in significant ways due to the differing contexts in which these skills were discussed. Table I shows frequency of mentions varies, with traditional software engineering skills being highlighted more often in job postings, while innovation and social responsibility are more prominent in practitioner interviews. This might be reflecting the differences in the perspectives of employers and employees, e.g., while employers are focused on ensuring general team efficiency and project success, practitioners are increasingly aware that addressing AI's ethical and societal challenges is a core part of their role. Yet, further investigations are necessary to claim this finding. Table II summarizes the soft skills identified and presents evidence extracted from the job postings and interviews.

Table I
NUMBER OF MENTIONS OF EACH SOFT SKILL

| Soft Skill | Number of Mentions |
| --- | --- |
| Clear communication | 93 |
| Problem-solving | 36 |
| Effective Collaboration | 35 |
| Leadership | 24 |
| Interpersonal Relationship | 14 |
| Decision-making | 11 |
| Empathy | 7 |
| Critical thinking | 5 |
| Teamwork | 5 |
| Time management | 5 |
| Curiosity | 3 |
| Ethical Awareness | 2 |

Job postings typically focus on general software engineering skills, such as coordination and management, which are broadly applicable to any role in software teams, not just data scientists. This focus explains why there were more mentions of these skills in job descriptions—they are essential for ensuring team efficiency, communication, and project management, which are critical to any successful software development process. In this context, employers prioritize skills that facilitate teamwork, leadership, and execution since these are foundational for any role involved in the lifecycle of software products. These broad skills are necessary for managing projects at scale and for ensuring that teams function smoothly, regardless of the specific technical challenges they face.

However, the interviews with practitioners provided a more focused lens on the specific demands of data science, particularly those associated with AI development. Practitioners, having direct experience with the technical and ethical challenges of AI, emphasized skills that are more closely related to identifying and mitigating AI-related issues, such as bias and ethical concerns. For these professionals, the ability to understand and address the nuances of data and algorithmic biases was not just an additional competency—it was a core aspect of their role. This difference highlights the evolving nature of the data scientist's role in AI-driven projects, where technical proficiency must be balanced with ethical awareness and a proactive approach to problem-solving.

## V. DISCUSSIONS

Soft skills have gained significant recognition in recent decades [64]. During the Fordism period, the primary focus was on manual skills, with little emphasis on soft skills [65]. However, the modern job market has shifted toward creative roles [66], with positions like software engineers, and data scientists becoming some of the most sought-after [67], [68]. In these roles, soft skills have become just as essential as technical expertise—if not more so, particularly when working directly with people [69]. Collaboration, for instance, is essential for project success, requiring team members to work together to achieve a common goal [70]. Furthermore, interacting with stakeholders to understand their needs and challenges is key to delivering effective solutions [71].

In software development, soft skills are particularly important for ensuring that solutions meet the needs of the target audience and are not based on flawed assumptions [72]. With the rapid advancement of AI and the growing challenge of identifying and mitigating bias or prejudice in AI systems, awareness of this problem and a commitment to minimizing its influence on society have become critical tasks for software professionals [73]. As a result, based on the evidence we collected in this study, data scientists must possess strong, soft skills related to identifying and addressing bias. Whether they are training deep learning models or making decisions based on data insights, data scientists must ensure that the outcomes of their work are fair, equitable, and free from biases related to race, gender, sexuality, religion, or other sensitive traits. These soft skills are essential to creating ethical and efficient AI solutions.

Our findings reveal that, beyond the common soft skills found in software engineering, two sets of skills are emerging as particularly important for modern data scientists: *innovation* and *social responsibility*. While traditional software engineering skills remain fundamental, the growing complexity and societal impact of AI systems are bringing these two areas to the forefront. *Innovation skills*, such as curiosity and critical thinking, enable data scientists to navigate the rapidly evolving AI landscape and address its unique challenges.

Equally important is the increasing emphasis on *social responsibility*. As AI becomes more embedded in everyday life, data scientists are expected to incorporate ethical considerations into their work. Skills like *empathy* and *ethical awareness* empower data scientists to critically evaluate the potential impacts of AI on users and society at large. This shift highlights that data scientists must not only ensure the technical accuracy of their models but also prioritize fairness, inclusivity, and broader societal implications.

TABLE II
SUBCATEGORY DEFINITIONS AND ILLUSTRATIVE QUOTATIONS*

| Category | Subcategory | Definition | Illustrative Quotation |
|---|---|---|---|
| Coordination | Effective collaboration | Seeks opportunities to work with diverse individuals and organizations [54]. | "Collaboration: work closely with data scientists, software engineers, and product teams to translate business requirements into technical solutions." (J7) |
| | Clear communication | Any act by which one person gives to or receives from another person information about that person's needs, desires, perceptions, knowledge, or affective states [55]. | "In terms of soft skills, the basics would be communication, since you're working in a team, so one person doesn't stay focused on just one thing, isolated from the rest of the team, not knowing what's going on." (P9) |
| | Teamwork | A joint action by a group of people, in which individual interests are subordinated to group unity and efficiency [56]. | "We are a united team that strongly believes in teamwork: together, we work, discuss, and achieve results that inspire us to reach increasingly ambitious goals." (J24) |
| | Interpersonal relationships | Involves relations between people [57]. | "Good interpersonal skills and the ability to quickly build effective working relationships with others." (J85) |
| Engineering | Decision Making | Makes effective, timely decisions informed by sound reasoning. Evaluates and assimilates information from multiple sources and considers the consequences of potential actions [54]. | "Implementing data analytics solutions to support business decision-making processes." (J32) |
| | Problem Solving | Evaluates options and selects appropriate actions when confronted with a problem. Identifies and fills gaps in information required. Applies analytic methods in solving problems [54]. | "Fantastic problem-solving and analytical skills, with the ability to translate complex technical details into understandable business insights." (J18) |
| Management | Leadership | The process of influencing the activities of an individual or group to achieve certain objectives in a given situation [58]. | "Demonstrating proactive leadership and taking full ownership of technical outcomes to ensure project success." (J74) |
| | Time management | Behaviours that aim at achieving an effective use of time while performing certain goal-directed activities [59]. | "Good time management with the ability to multitask, prioritise and deliver to schedule." (J87) |
| Innovation | Curiosity | Curiosity is a durable individual difference or personality trait that prescribes individuals' typical exploratory responses occurring as part of dynamic states [63]. | "Right, I think the main thing is curiosity, because it's curiosity that will make us investigate data problems more thoroughly and also want to understand business rules." (P6) "I think it would be interesting to have very curious people dealing with data issues to seek a better understanding of it, so they can perform better processing and perhaps choose a model that fits the data set better." (P11) |
| | Critical Thinking | The ability to think is focused on the breakdown of what must be believed or done to make decisions in resolving problems [60]. | "Strong candidates will exhibit solid critical thinking skills, the ability to synthesize complex problems, and a talent for transforming data to create solutions that add value to a myriad of business requirements." (J46) "Analytical thinking: strong analytical and critical thinking abilities." (J62) |
| Social Responsability | Empathy | Empathy indicates an emotional rapport, or identification, with another person [61]. | "Yes, I don't know if I would call it empathy, but I think it would be a soft skill that some people need to develop." (P1) "We thrive on customer empathy, which remains our focus when creating excellent customer experiences through product innovation." (J14) |
| | Ethical Awareness | The discipline dealing with what is good and bad and with moral duty and obligation [62]. | "There should be more subjects focused on ethics, especially from an educational perspective, because we're dealing with data, and data is a sensitive thing, right?" (P3) "...we are a company that rewards initiative, resourcefulness, and work ethic." (J40) |

*more supporting evidence are available at https://figshare.com/s/4fda5ee8e36e1482db61

### A. Implications to Research

Our findings contribute significantly to the body of knowledge about soft skills in software engineering, particularly in the context of AI-driven solutions. Previous studies have typically focused on general soft skills necessary for collaboration and teamwork in software development. However, our work extends this by identifying soft skills that are becoming increasingly relevant in the AI era, such as *empathy* and *ethical awareness*. These skills highlight the evolving demands placed on data scientists and software professionals, emphasizing the need for solutions that account for societal impact. By incorporating these skills into discussions about software engineering, we add depth to the understanding of how technical roles are expanding to include socio-ethical responsibilities.

Moreover, our study contributes to the ongoing conversation around the necessity of socio-technical approaches in AI development. The inclusion of empathy and ethical awareness in our findings demonstrates the growing importance of aligning technical expertise with social consciousness. Professionals working with AI development, e.g., data scientists, are no longer expected to simply deliver functional systems; they are also expected to consider the broader implications of their work on society. This shift calls for more integrated approaches that blend technical precision with an understanding of human needs and ethical considerations. Our findings provide concrete evidence supporting the need for such approaches in AI development.

In terms of opportunities for future research, our work opens several avenues. First, the relatively recent emergence of empathy and ethical awareness as relevant skills suggests that these areas warrant further investigation. Research could explore how these skills are developed and measured in software professionals, particularly in relation to mitigating bias in AI systems. Additionally, the disparity between the soft skills emphasized in job postings and those highlighted in practitioner interviews suggests a potential gap in how the industry perceives the role of soft skills versus how they are applied in practice. Future studies could examine this disconnect, providing insights into how hiring practices and training programs could be better aligned with the evolving needs of the software industry.

Finally, by identifying empathy and ethical awareness as emerging soft skills, we encourage future research to explore how these competencies influence the design and implementation of AI systems. This could include examining the impact of these skills on reducing bias in machine learning models, improving user-centric design, and fostering fairer, more inclusive AI systems. These research opportunities are important for ensuring that AI technologies not only meet technical specifications but also address the ethical and societal challenges that come with their increasing use in critical decision-making.

### B. Implications to Practice

For practitioners, particularly data scientists, our findings offer valuable insights into the importance of soft skills in AI-driven solutions. As the AI field continues to evolve, the ability to identify and mitigate bias, ensure fairness, and consider broader societal impacts is becoming a critical aspect of their work. Professionals who cultivate strong soft skills—such as *empathy* and *ethical awareness*—will be better equipped to design AI systems that respect privacy, fairness, and human dignity. By developing these skills, data scientists can position themselves as leaders in the creation of AI solutions that not only function effectively but also contribute positively to society.

Additionally, software companies can use our insights to shape their recruitment strategies and organizational culture. By recognizing the emerging importance of soft skills like *curiosity*, *critical thinking*, *empathy*, and *ethical awareness*, companies can successfully identify professionals who demonstrate these competencies alongside technical proficiency. This will enable them to anticipate and address ethical and social challenges in AI development, positioning themselves at the forefront of the industry. Additionally, companies can implement internal training programs to foster these skills within their existing teams. Creating a culture that emphasizes user-oriented development—where diversity, equity, fairness, and inclusivity are prioritized will allow companies to produce socially responsible AI systems. This can be achieved by creating campaigns that foster awareness about specific topics, integrating people from minority groups into development teams, and ensuring feedback is collected on their solutions from a wide range of users with diverse backgrounds. Such approaches can improve the reputation, build customer trust, and drive revenue growth.

Finally, we must highlight the broader impact of our findings on AI development. As AI becomes increasingly embedded in decision-making processes across industries, the need for ethically sound, user-centric AI systems is more pressing than ever. Our research underscores the growing importance of soft skills in mitigating bias, promoting inclusivity, and safeguarding against harmful AI outcomes. By emphasizing key competencies such as *curiosity*, *critical thinking*, *empathy*, and *ethical awareness*, alongside other general skills related to *coordination*, *engineering*, and *management*, our findings advocate for a shift toward more responsible AI development practices. Ultimately, this approach can foster greater trust between AI developers and the communities affected by these technologies, ensuring that AI systems not only function effectively but also align with societal values.

### C. Threats to Validity

From the beginning, we aimed to collect data that reflected both the industry's and practitioners' perspectives to provide a more comprehensive and generalized analysis. However, despite our efforts, our method has some limitations that should be acknowledged. For example, one key limitation of our study is the focus on English-language job postings, which restricted our search to English-speaking markets and excluded non-English postings where different terms or nuances may describe similar soft skills. This could have caused us to miss important skills in non-English-speaking countries, limiting the global relevance of our findings. Additionally, regional differences in job posting platforms further narrowed our data. While LinkedIn is common in Western countries, many regions use local platforms that we did not capture. As a result, our dataset may not fully reflect the global landscape of soft skills, suggesting the need for a broader international perspective.

In terms of our interviews, a further limitation was the geographic and cultural homogeneity of our participants. Most of the professionals we interviewed were working for Western companies, which limited the diversity of perspectives we could gather. This lack of geographic diversity meant that we missed the opportunity to explore how professionals in non-Western contexts view soft skills. Different cultural values, organizational practices, and regional industry standards may influence how soft skills are perceived and applied in various contexts, making our findings more reflective of Western practices and less representative of global trends.

Moreover, the nature of our method, coupled with the relatively small sample size, restricts our ability to make statistical generalizations to a broader population. Therefore, we do not claim statistical generalizations. However, consistent with the nature of qualitative research, while our findings provide valuable insights, they are best understood as *transferable* to different contexts using analytical strategies. In other words, our conclusions can be applied to comparable studies or contexts but should not be assumed to apply across all settings. Despite these limitations, our research provides insights about industry requirements and practitioner experiences, offering a comprehensive view of the evolving soft skills landscape within AI development.

Finally, following the principles of qualitative research, it is important to recognize the balance between the contextual limitations of our study and the broader applicability of our findings. In this regard, our study makes a meaningful contribution to the broader body of knowledge by exploring the intersection of traditional soft skills with emerging trends in AI and data science. We highlight the growing importance of *empathy*, *ethical awareness*, and *innovation* as critical components in the development of AI-driven solutions. These insights have implications for both industry and academia, suggesting new avenues for research and practical applications that focus on integrating socio-technical skills into AI development.

## VI. CONCLUSION

In this study, we conducted a qualitative analysis using thematic analysis to identify the essential soft skills required in AI-related roles. We collected data from job postings and interviews with data scientists working on AI systems projects. Our findings revealed that the most highly valued soft skills in the industry today focus on *coordination*, *engineering*, *management*, *innovation*, and *social responsibility*. Notably, there is a growing emphasis on emerging soft skills such as *empathy* and *ethical awareness*, particularly in projects involving AI and machine learning models. These skills are becoming increasingly relevant, as AI solutions have the potential to impact society in significant ways. Data scientists are now expected to make key decisions related to bias and fairness, ensuring that AI systems operate equitably and without bias, particularly in how they handle and process data.

In conclusion, integrating social skills such as *empathy* and *ethical awareness* into AI-driven projects presents significant opportunities for both organizations and professionals. As AI systems increasingly influence decision-making across industries, the need for a balanced combination of technical and socio-technical skills continues to grow. Software companies that prioritize these skills in their processes are better positioned to develop responsible, user-centered AI systems that uphold privacy, fairness, and human dignity—ultimately fostering public trust and achieving long-term success. For professionals, honing these skills not only enhances their ability to deliver more effective and ethical AI solutions but also boosts their competitiveness in the evolving job market. We have gathered significant insights from this rapidly evolving field by drawing on data from job postings and practitioners across 12 countries and six regions worldwide. Despite the limited sample size, this qualitative study aimed to identify the most sought-after data skills required to develop AI systems. Our research initiates a discussion on this topic by presenting transferable findings that, as explained in the threats to validity section, are not universally generalizable. This was not our focus, as this is a qualitative study, and we did not seek statistical generalization. However, consistent with our method, we expect that our findings can be re-analyzed and transferred to different contexts in an analytical way. Future work may focus on statistical methods and draw on larger samples to better understand how these soft skills play a role in larger populations of professionals around the world. As the AI industry continues to expand, those who prioritize *empathy*, *ethical awareness*, and socio-technical responsibility will be well-equipped to lead the development of AI technologies that serve society in a responsible and equitable manner.

## VII. DATA AVAILABILITY

Job postings and a sample of safe quotations extracted from the interviews are available at: https://figshare.com/s/4fda5ee8e36e1482db61. We removed any quotations that included direct references to participants' names, co-workers, projects, or companies. As a result, some participants from our sample may not have associated quotations in the spreadsheet. Additionally, some quotations might read awkwardly as they were directly translated from the participants' native languages.

## REFERENCES

[1] E. M. Dennis, "Project success factors when implementing and maintaining information technologies," Ph.D. dissertation, Colorado Technical University, 2015.

[2] K. Albusays, P. Bjorn, L. Dabbish, D. Ford, E. Murphy-Hill, A. Serebrenik, and M.-A. Storey, "The diversity crisis in software development," *IEEE Software*, vol. 38, no. 2, pp. 19–25, 2021.

[3] S. Hyrynsalmi, M. M. Rantanen, and S. Hyrynsalmi, "The skill gap and polarization of the software labour force: Early signs of the war of talents between software professionals and how it threatens wellbeing," *Finnish Journal of eHealth and eWelfare*, vol. 13, no. 2, pp. 113–123, 2021.

[4] H. Karimi and A. Pina, "Strategically addressing the soft skills gap among stem undergraduates," *Journal of Research in STEM Education*, vol. 7, no. 1, pp. 21–46, 2021.

[5] K. Flaherty, "Soft skills: The critical accompaniment to technical skills." *AMWA Journal: American Medical Writers Association Journal*, vol. 29, no. 2, 2014.

[6] E. Rainsbury, D. L. Hodges, N. Burchell, and M. C. Lay, "Ranking workplace competencies: Student and graduate perceptions," 2002.

[7] A. Gibert, W. C. Tozer, and M. Westoby, "Teamwork, soft skills, and research training," *Trends in ecology & evolution*, vol. 32, no. 2, pp. 81–84, 2017.

[8] M. Omar, M. Rehman, and A. Amin, "Does software requirement elicitation and personality make any relation?" 2019.

[9] C. Wohlin, A. Aurum, L. Angelis, L. Phillips, Y. Dittrich, T. Gorschek, H. Grahn, K. Henningsson, S. Kagstrom, G. Low et al., "The success factors powering industry-academia collaboration," *IEEE software*, vol. 29, no. 2, pp. 67–73, 2011.

[10] Y. Lindsjørn, D. I. Sjøberg, T. Dingsøyr, G. R. Bergersen, and T. Dybå, "Teamwork quality and project success in software development: A survey of agile development teams," *Journal of Systems and Software*, vol. 122, pp. 274–286, 2016.

[11] H. K. Dam, "Artificial intelligence for software engineering," *XRDS: Crossroads, The ACM Magazine for Students*, vol. 25, no. 3, pp. 34–37, 2019.

[12] Z. Wan, X. Xia, D. Lo, and G. C. Murphy, "How does machine learning change software development practices?" *IEEE Transactions on Software Engineering*, vol. 47, no. 9, pp. 1857–1871, 2019.

[13] J. Sauvola, S. Tarkoma, M. Klemettinen, J. Riekki, and D. Doermann, "Future of software development with generative ai," *Automated Software Engineering*, vol. 31, no. 1, p. 26, 2024.

[14] A. Min, "Artificial intelligence and bias: Challenges, implications, and remedies." *Journal of Social Research*, vol. 2, no. 11, 2023.

[15] N. Mehrabi, F. Morstatter, N. Saxena, K. Lerman, and A. Galstyan, "A survey on bias and fairness in machine learning," *ACM computing surveys (CSUR)*, vol. 54, no. 6, pp. 1–35, 2021.

[16] C. Huang, Z. Zhang, B. Mao, and X. Yao, "An overview of artificial intelligence ethics," *IEEE Transactions on Artificial Intelligence*, vol. 4, no. 4, pp. 799–819, 2022.

[17] N.-J. Iloanusi and S. A. Chun, "Ai impact on health equity for marginalized, racial, and ethnic minorities," in *Proceedings of the 25th Annual International Conference on Digital Government Research*, 2024, pp. 841–848.

[18] A. Caliskan, "Artificial intelligence, bias, and ethics." in *IJCAI*, 2023, pp. 7007–7013.

[19] X.-W. Chen and X. Lin, "Big data deep learning: challenges and perspectives," *IEEE access*, vol. 2, pp. 514–525, 2014.

[20] J. M. Nolin, "Data as oil, infrastructure or asset? three metaphors of data as economic value," *Journal of Information, Communication and Ethics in Society*, vol. 18, no. 1, pp. 28–43, 2020.

[21] A. Ahmed and A. M. Abdulkareem, "Big data analytics in the entertainment industry: audience behavior analysis, content recommendation, and revenue maximization," *Reviews of Contemporary Business Analytics*, vol. 6, no. 1, pp. 88–102, 2023.

[22] X. Amatriain, "Big & personal: data and models behind netflix recommendations," in *Proceedings of the 2nd international workshop on big data, streams and heterogeneous source Mining: Algorithms, systems, programming models and applications*, 2013, pp. 1–6.

[23] C. Barroso-Moreno, L. Rayon-Rumayor, and A. B. García-Vera, "Big data and business intelligence on twitter and instagram for digital inclusion." *Comunicar: Media Education Research Journal*, vol. 31, no. 74, pp. 45–56, 2023.

[24] K. Gauen, R. Dailey, J. Laiman, Y. Zi, N. Asokan, Y.-H. Lu, G. K. Thiruvathukal, M.-L. Shyu, and S.-C. Chen, "Comparison of visual datasets for machine learning," in *2017 IEEE International Conference on Information Reuse and Integration (IRI)*. IEEE, 2017, pp. 346–355.

[25] T. H. Davenport and D. Patil, "Data scientist," *Harvard business review*, vol. 90, no. 5, pp. 70–76, 2012.

[26] J. L. Correia, J. A. Pereira, R. Mello, A. Garcia, B. Fonseca, M. Ribeiro, R. Gheyi, M. Kalinowski, R. Cerqueira, and W. Tiengo, "Brazilian data scientists: revealing their challenges and practices on machine learning model development," in *Proceedings of the XIX Brazilian Symposium on Software Quality*, 2020, pp. 1–10.

[27] C. C. da Silveira, C. B. Marcolin, M. da Silva, and J. C. Domingos, "What is a data scientist? analysis of core soft and technical competencies in job postings," *Revista Inovação, Projetos e Tecnologias*, vol. 8, no. 1, pp. 25–39, 2020.

[28] G. Matturro, F. Raschetti, and C. Fontán, "A systematic mapping study on soft skills in software engineering." *J. Univers. Comput. Sci.*, vol. 25, no. 1, pp. 16–41, 2019.

[29] G. Matturro, "Soft skills in software engineering: A study of its demand by software companies in uruguay," in *2013 6th international workshop on cooperative and human aspects of software engineering (CHASE)*. IEEE, 2013, pp. 133–136.

[30] F. Ahmed, L. F. Capretz, S. Bouktif, and P. Campbell, "Soft skills and software development: A reflection from the software industry," *arXiv preprint arXiv:1507.06873*, 2015.

[31] M. Hancher, "Humpty dumpty and verbal meaning," *The Journal of Aesthetics and Art Criticism*, vol. 40, no. 1, pp. 49–58, 1981.

[32] A. Hidayati, E. K. Budiardjo, and B. Purwandari, "Hard and soft skills for scrum global software development teams," in *Proceedings of the 3rd International Conference on Software Engineering and Information Management*, 2020, pp. 110–114.

[33] M. L. Matteson, L. Anderson, and C. Boyden, "" soft skills": A phrase in search of meaning," *portal: Libraries and the Academy*, vol. 16, no. 1, pp. 71–88, 2016.

[34] F. Ahmed, L. F. Capretz, and P. Campbell, "Evaluating the demand for soft skills in software development," *It Professional*, vol. 14, no. 1, pp. 44–49, 2012.

[35] A. Sukhoo, A. Barnard, M. M. Eloff, J. A. Van der Poll, and M. Motah, "Accommodating soft skills in software project management." *Issues in Informing Science & Information Technology*, vol. 2, 2005.

[36] D. B. de Campos, L. M. M. de Resende, and A. B. Fagundes, "The importance of soft skills for the engineering," *Creative Education*, vol. 11, no. 8, pp. 1504–1520, 2020.

[37] G. Matturro, F. Raschetti, and C. Fontán, "Soft skills in software development teams: A survey of the points of view of team leaders and team members," in *2015 IEEE/ACM 8th International Workshop on Cooperative and Human Aspects of Software Engineering*. IEEE, 2015, pp. 101–104.

[38] W. Groeneveld, L. Luyten, J. Vennekens, and K. Aerts, "Exploring the role of creativity in software engineering," in *2021 IEEE/ACM 43rd International Conference on Software Engineering: Software Engineering in Society (ICSE-SEIS)*. IEEE, 2021, pp. 1–9.

[39] A. Meier, M. Kropp, and G. Perellano, "Experience report of teaching agile collaboration and values: agile software development in large student teams," in *2016 IEEE 29th International Conference on Software Engineering Education and Training (CSEET)*. IEEE, 2016, pp. 76–80.

[40] I. Ilavarasi, "Enhancing workplace productivity: A review of effective communication techniques and their role in fostering team collaboration and conflict resolution," *International Journal for Multidimensional Research Perspectives*, vol. 2, no. 4, pp. 33–45, 2024.

[41] M. Korkala, P. Abrahamsson, and P. Kyllonen, "A case study on the impact of customer communication on defects in agile software development," in *AGILE 2006 (AGILE'06)*. IEEE, 2006, pp. 11–pp.

[42] N. C. Pa and A. M. Zin, "Requirement elicitation: identifying the communication challenges between developer and customer," *International Journal of New Computer Architectures and their Applications (IJNCAA)*, vol. 1, no. 2, pp. 371–383, 2011.

[43] C. V. Magalhães, F. Q. da Silva, and R. E. Santos, "The role of job specialization in the software industry," in *International Conference on Information Technology & Systems*. Springer, 2022, pp. 307–317.

[44] N. M. Agrawal and M. Thite, "Nature and importance of soft skills in software project leaders," *IIM Bangalore Research Paper*, no. 214, 2003.

[45] R. Florea and V. Stray, "Software tester, we want to hire you! an analysis of the demand for soft skills," in *Agile Processes in Software Engineering and Extreme Programming: 19th International Conference, XP 2018, Porto, Portugal, May 21–25, 2018, Proceedings 19*. Springer, 2018, pp. 54–67.

[46] J. A. Maxwell *et al.*, *Designing a qualitative study*. The SAGE handbook of applied social research methods, 2008, vol. 2.

[47] J. Thielen and A. Neeser, "Making job postings more equitable: Evidence based recommendations from an analysis of data professionals job postings between 2013-2018," *Evidence Based Library and Information Practice*, vol. 15, no. 3, pp. 103–156, 2020.

[48] H. Alshenqeeti, "Interviewing as a data collection method: A critical review," *English linguistics research*, vol. 3, no. 1, pp. 39–45, 2014.

[49] Apr 2024. [Online]. Available: https://www.imf.org/en/Publications/WEO/weo-database/2024/April

[50] S. Baltes and P. Ralph, "Sampling in software engineering research: A critical review and guidelines," *Empirical Software Engineering*, vol. 27, no. 4, pp. 1–31, 2022.

[51] K. Charmaz, *Constructing grounded theory*. sage, 2014.

[52] D. S. Cruzes and T. Dyba, "Recommended steps for thematic synthesis in software engineering," in *2011 international symposium on empirical software engineering and measurement*. IEEE, 2011, pp. 275–284.

[53] T. Heston and C. Khun, "Prompt engineering in medical education," *International Medical Education*, vol. 2, pp. 198–205, 8 2023.

[54] L. G. Barron and M. R. Rose, "Malleability of soft-skill competencies," *military learning*, p. 3, 2021.

[55] J. Velentzas and G. Broni, "Communication cycle: Definition, process, models and examples," *Recent advances in financial planning and product development*, vol. 17, pp. 117–131, 2014.

[56] R. Peter and M. S. Simatupang, "Teamwork soft skill development in facing the globalization," *International Journal of Engineering and Advanced Technology (IJEAT)*, vol. 8, pp. 486–491, 2019.

[57] A. G. Mokoena-De Beer, "Facilitation of constructive intra-and interpersonal relationships: A concept analysis," *Health SA Gesondheid*, vol. 29, p. 8, 2024.

[58] R. A. Barker, "How can we train leaders if we do not know what leadership is?" *Human relations*, vol. 50, no. 4, pp. 343–362, 1997.

[59] B. J. Claessens, W. Van Eerde, C. G. Rutte, and R. A. Roe, "A review of the time management literature," *Personnel review*, vol. 36, no. 2, pp. 255–276, 2007.

[60] L. Faridah and A. Nasikhah, "Improve critical thinking ability and mathematical representation of junior high school students throught softskill based metacognitive approaches," in *International Conference on Science, Technology, Education, Arts, Culture and Humanity-" Interdisciplinary Challenges for Humanity Education in Digital Era"(STEACH 2018)*. Atlantis Press, 2019, pp. 73–76.

[61] E. Lord-Kambitsch, "Introduction to empathy: Activation, definition, construct," *Think Pieces: A Journal of the Arts, Humanities, and Social Sciences*, vol. 1, no. 1, pp. 1–8, 2014.

[62] C. L. Derr, "Ethics and leadership," *Journal of Leadership, accountability and ethics*, vol. 9, no. 6, pp. 66–71, 2012.

[63] A. Horstmeyer, "The generative role of curiosity in soft skills development for contemporary vuca environments," *Journal of Organizational Change Management*, vol. 33, no. 5, pp. 737–751, 2020.

[64] C. Succi and M. Canovi, "Soft skills to enhance graduate employability: comparing students and employers' perceptions," *Studies in higher education*, vol. 45, no. 9, pp. 1834–1847, 2020.

[65] A. Sorgner, "Technological development, changes on labor markets, and demand for skills," *Форсайт*, vol. 13, no. 2 (eng), pp. 6–8, 2019.

[66] A. K. Rajeev, "Contribution of small and medium creative enterprises to the national economy: A systematic review," in *Turiba University. International Scientific Conference*. Turiba University, 2022, pp. 98–106.

[67] C. T. Oyewale, A. M. Abolade, and O. Oyewumi, "Effective working environment and factor for a software engineer in companies that are not ict based," in *Proceedings of the 6th ACM SIGCAS/SIGCHI Conference on Computing and Sustainable Societies*, 2023, pp. 155–158.

[68] N. Bhalli, V. Janeja, and D. Harding, "Effects of prior academic experience in introductory level data science course," in *Proceedings of the 55th ACM Technical Symposium on Computer Science Education V. 2*, 2024, pp. 1576–1577.

[69] L. F. Capretz and F. Ahmed, "A call to promote soft skills in software engineering," *arXiv preprint arXiv:1901.01819*, 2018.

[70] N. Oza, F. Fagerholm, and J. Münch, "How does kanban impact communication and collaboration in software engineering teams?" in *2013 6th International Workshop on Cooperative and Human Aspects of Software Engineering (CHASE)*. IEEE, 2013, pp. 125–128.

[71] L. R. Wong, D. S. Mauricio, G. D. Rodriguez *et al.*, "A systematic literature review about software requirements elicitation," *Journal of Engineering Science and Technology*, vol. 12, no. 2, pp. 296–317, 2017.

[72] E. Köppen, I. Rauth, M. Schnjakin, and C. Meinel, "The importance of empathy in it projects: a case study on the development of the german electronic identity card," in *DS 68-7: Proceedings of the 18th International Conference on Engineering Design (ICED 11), Impacting Society through Engineering Design, Vol. 7: Human Behaviour in Design, Lyngby/Copenhagen, Denmark, 15.-19.08. 2011*, 2011.

[73] M. DeCamp and C. Lindvall, "Mitigating bias in ai at the point of care," *Science*, vol. 381, no. 6654, pp. 150–152, 2023.